\documentclass[aps,prl,twocolumn]{revtex4}
\usepackage{epsfig}
\usepackage{graphicx}
\usepackage{bm}
\begin{document}

\title{Photon-photon gates in Bose-Einstein condensates}

\author{Arnaud Rispe, Bing He, and Christoph Simon}
\affiliation{Institute for Quantum Information Science and
Department of Physics and Astronomy, University of Calgary,
Calgary T2N 1N4, Alberta, Canada}

\begin{abstract}
It has recently been shown that light can be stored in
Bose-Einstein condensates for over a second. Here we
propose a method for realizing a controlled phase gate
between two stored photons. The photons are both stored in
the ground state of the effective trapping potential inside
the condensate. The collision-induced interaction is
enhanced by adiabatically increasing the trapping frequency
and by using a Feshbach resonance. A controlled phase shift
of $\pi$ can be achieved in one second.
\end{abstract}
\date{\today}

\maketitle

Photons are ideal carriers of quantum information over long
distances. It is interesting to explore their potential for
the implementation of quantum information processing as
well. This is particularly relevant for quantum repeaters
\cite{briegel,dlcz,sangouard}, which would allow one to
distribute quantum states over distances that are
inaccessible by direct transmission. Quantum repeaters
require both the capability to store photons for relatively long times and to perform
efficient quantum gates between them \cite{sangouard}. Potential
architectures where storage and quantum gates can be
achieved in the same system are particularly attractive.
Recently it was shown that light can be stored for over a
second in a Bose-Einstein condensate (BEC) \cite{zhang},
making condensates a very interesting candidate system for
the implementation of quantum memories. Quantum repeaters can
tolerate long gate times in the sub-second range, since
repetition rates are in any case limited by other factors
such as communication times and transmission probabilities.
It is therefore of great interest to explore the potential
for photon-photon gates in BECs, where interactions between
stored excitations are weak, but non-zero.

In the following we describe a concrete proposal for
realizing such photon-photon gates in BECs. Our work builds
on Refs. \cite{duttonhau,duttonclark}, but we focus on the
case of two single photons interacting. In this extremely
low-intensity regime achieving significant controlled phase
shifts is not straightforward. However, we show that phase
shifts of $\pi$ can be achieved on sub-second timescales by
combining a Feshbach enhancement of the relevant
scattering length and an adiabatic
compression of the trap after the light has been stored.
The fidelity of photon-photon gates can be affected by
unwanted multi-mode effects, see e.g. Ref. \cite{he}. In
the present proposal these effects are greatly suppressed
by the fact that the interaction is much weaker than the
confinement, ensuring high-fidelity operations.

\begin{figure}
\epsfig{file=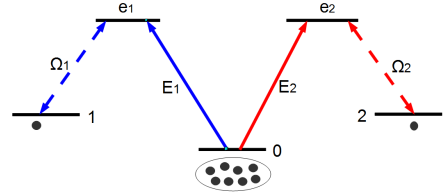,width=0.7 \columnwidth}
\caption{Level scheme for photon-photon gate. The BEC is
prepared in level 0. The single photons in modes $E_1$ and
$E_2$ can be independently stored as delocalized
excitations in levels 1 and 2, using the control beams
$\Omega_1$ and $\Omega_2$.} \label{levels}
\end{figure}

Let us assume that the two photons have orthogonal
polarization. Their propagation inside the BEC can be
controlled by two independent control beams, leading to
storage in two different atomic levels 1 and 2, where the
BEC was prepared in level 0, see Fig. \ref{levels}. Slow
and stopped light in BECs has been thoroughly investigated
\cite{zhang,duttonhau,hauslow,haumemory,ginsberg}.
Due to the linearity of the equations of motion, the physics of storage and retrieval is the same at the
single-photon level as for weak classical probe pulses
\cite{gorshkov,hammerer}. Inside the medium and in the
presence of the appropriate control beam, the photon is
converted into a slowly moving polariton, which can be
stopped by adiabatically switching off the control beam,
thus converting the photon into a stored atomic spin wave.
Running the process in reverse allows the reconversion of
spin waves into photons. Here we focus on the interaction
between the two spin waves, once the control beams have
been turned off. Due to the weakness of the
collision-induced interactions the timescale for the
storage and retrieval processes is much shorter than the
timescale on which significant interaction occurs in the
photon-photon regime.

The dynamics of the atomic spin waves is governed by the
collisional interactions between atoms in combination with
the external trapping potential. Spin waves in levels 1 and
2 experience an effective trapping potential and effective
collisional interactions that depend on the differences
between the atomic scattering lengths in the various atomic
levels \cite{duttonclark}. These differences, which are
usually small, can be enhanced by Feshbach resonances
\cite{inouye,volz,kaufman}. We consider a situation where
both spin waves experience the same effective trapping
potential, and where they are both in its ground state. The
latter condition can be achieved by carefully matching the
pulse duration and width of the incoming photons and the
intensity of the control beams (which determines the
propagation speed and thus the longitudinal extent of the
polaritons inside the condensate) to the parameters of the
effective trapping potential. We focus on the regime where the stored spin waves are localized well inside the condensate, cf. Fig. 2.

\begin{figure}
\epsfig{file=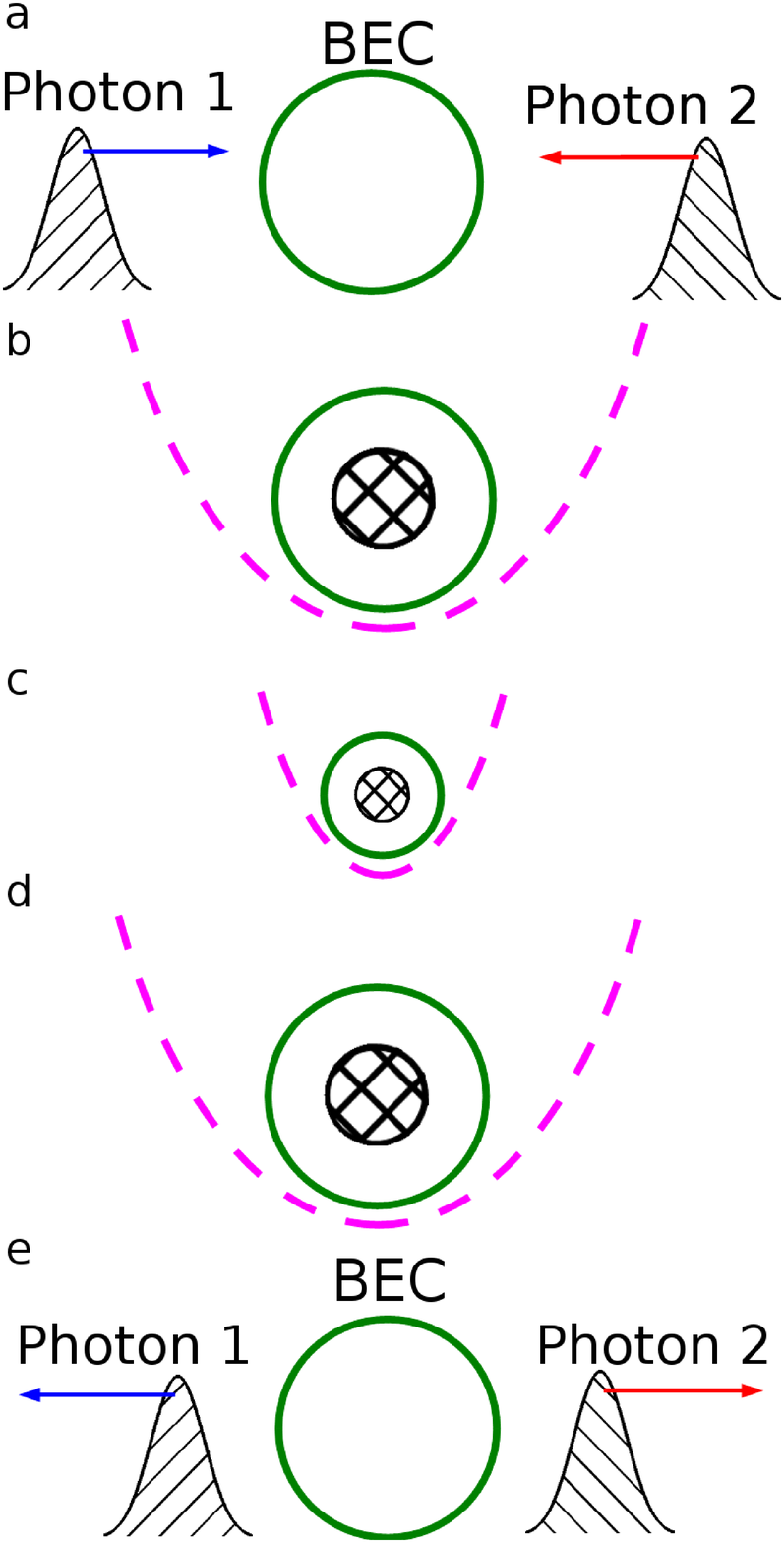,width=0.5\columnwidth}
\caption{Principle of photon-photon gate. (a) Photons 1 and
2 are independently absorbed by the BEC. (b) Both are
converted into atomic excitations that are in the ground
state of the effective trapping potential, see text. (c) The
collision-induced interaction between the atomic spin waves
is enhanced by adiabatically increasing the trapping
potential, thus reducing the size of the ground state wave
functions (and of the BEC). (d) The trapping potential is
adiabatically brought back to its original value. (e) The
photons can be read out independently.} \label{gate}
\end{figure}

The interaction strength, and thus the accumulated
controlled phase shift for a given time, then strongly
depends not only on the scattering lengths, but also on the
size of the ground state wave packets. During storage and
retrieval, this size has to be significantly larger than a
wavelength, due to focusing restrictions for the transverse
dimensions and in order to justify the slowly varying
envelope description which underlies the polariton picture
for the longitudinal dimension. However, in between storage
and retrieval it is possible to adiabatically increase the
trapping potential, thus reducing the size of the ground
state wave packets while keeping the spin waves in the
ground state, see Fig. \ref{gate}. This enhances
the interaction strength, making controlled phase shifts of
$\pi$ achievable on one-second timescales. Note that the
basic ingredients of the present proposal are similar to
those of single-atom quantum gates schemes based on cold
collisions such as Refs. \cite{jaksch,calarco}.

We now describe our proposal in more quantitative terms.
Our treatment of the spin waves inside the BEC is based on
Refs. \cite{duttonclark,duttonthesis}. The Gross-Pitaevskii
(GP) equation for the macroscopic wave-function $\psi_0$ of
the condensate is
\begin{eqnarray}
i\hbar \frac{\partial \psi_0}{\partial t}=(
-\frac{\hbar^2}{2m} \nabla^2 +V({\bf x})+U_{00}
|\psi_0|^2\nonumber\\ +U_{01} |\psi_1|^2 + U_{02}
|\psi_2|^2 ) \, \psi_0,
\end{eqnarray}
where $m$ is the atomic mass, $V$ is the trapping
potential, $U_{00}, U_{01}, U_{02}$ are the collisional
interaction potentials, which are related to the
corresponding scattering lengths $a_{00}, a_{01}, a_{02}$
by $U_{0j}=\frac{4\pi\hbar^2 a_{0j}}{m}$, and $\psi_1,
\psi_2$ are the macroscopic wave functions for levels 1 and
2. We will make the transition to a single-quantum
description for the latter in a moment.

For a sufficiently large condensate, and keeping in mind that the perturbation due to the spin waves in levels 1 and 2 is extremely weak in our case, the solution for $\psi_0$ will be essentially
stationary, and the stationary equation for a chemical
potential $\mu$ can be solved in the Thomas-Fermi
approximation (i.e. neglecting the kinetic term)
\cite{duttonthesis}, giving
\begin{equation}
|\psi_0|^2=\frac{1}{U_{00}}(\mu -V- U_{01} |\psi_1|^2 +
U_{02} |\psi_2|^2 ), \label{thomasfermi}
\end{equation}
where $\mu$ is the chemical potential.
 This solution of Eq. (\ref{thomasfermi}) can
now be inserted into the GP equations for $\psi_1$ and
$\psi_2$. Corrections to the Thomas-Fermi approximation mainly affect the boundary layer of the condensate \cite{TF}. We therefore expect the present treatment to be correct under the above-mentioned condition that the spin waves in levels 1 and 2 are localized well inside the BEC.

In order to describe the few-excitation regime, we replace the macroscopic wave
functions $\psi_1, \psi_2$ by quantum
field operators $\hat{\psi}_1, \hat{\psi}_2$ satisfying
commutation relations $[\hat{\psi}_i({\bf
x}),\hat{\psi}_j^{\dagger}({\bf x'})]=\delta_{ij}
\delta^{(3)}({\bf x-x'})$, in analogy to the transition
from classical to quantum non-linear optics \cite{hillery}. They fulfill
the equations (neglecting a constant energy shift that
depends on $\mu$)
\begin{eqnarray}
i\hbar \frac{\partial \hat{\psi}_1}{\partial t}=(
-\frac{\hbar^2}{2m} \nabla^2 +\tilde{V}_1({\bf
x})+\tilde{U}_{11} \hat{\psi}_1^\dagger \hat{\psi}_1
+\tilde{U}_{12} \hat{\psi}_2^\dagger \hat{\psi}_2 ) \,
\hat{\psi}_1 \nonumber\\
i\hbar \frac{\partial \hat{\psi}_2}{\partial t}=(
-\frac{\hbar^2}{2m} \nabla^2 +\tilde{V}_2({\bf
x})+\tilde{U}_{22} \hat{\psi}_2^\dagger \hat{\psi}_2
+\tilde{U}_{12} \hat{\psi}_1^\dagger \hat{\psi}_1 ) \,
\hat{\psi}_2, \label{qfields}
\end{eqnarray}
where $\tilde{V}_i=(1-\frac{a_{0i}}{a_{00}}) V$ are the
effective trapping potentials and
$\tilde{U}_{ij}=\frac{4\pi\hbar^2}{m}(a_{ij}-\frac{a_{0i}
a_{0j}}{a_{00}})$ are the effective interaction potentials,
which are all modified due to the interaction with the
background condensate. These equations are analogous to
those obtained in Ref. \cite{duttonclark} for the two-level
case. Here we have assumed that the bare trapping potential
$V$ is the same for all atomic levels. Moreover for
simplicity we will assume that $a_{01}=a_{02}$ implying
$\tilde{V}_1=\tilde{V}_2=\tilde{V}$. We require
$a_{01}<a_{00}$ in order for $\tilde{V}$ to be attractive,
provided that $V$ is attractive \cite{phasesep}.

Eq. (\ref{qfields}) allows one to describe the dynamics of
any number of quantum excitations in levels 1 and 2.
However, we are interested in the case where there is
exactly one excitation in each level. It is then convenient
to introduce the two-particle wave-function $\psi_{12}({\bf
x}_1,{\bf x}_2)=\langle 0|\hat{\psi}_1({\bf x}_1)
\hat{\psi}_2 ({\bf x}_2)|\Phi\rangle$, where $|0\rangle$ is
the state without any excitations (i.e. the state where
there is just the condensate in level 0), and
\begin{equation}
|\Phi\rangle=\int d^3 x_1 d^3 x_2 \phi_0({\bf x}_1)
\phi_0({\bf x}_2) \hat{\psi}^{\dagger}_1({\bf x}_1)
\hat{\psi}^{\dagger}_2 ({\bf x}_2)|0\rangle
\end{equation}
is the initial state (after storage), which consists of one
atomic excitation in each level (1 and 2), both of which
are in the ground state $\phi_0$ of the effective trapping
potential $\tilde{V}$. In the Heisenberg picture for the quantum field theory the state
remains constant, but the field operators evolve according
to Eq. (\ref{qfields}). As a consequence, the two-particle
wave function $\psi_{12}$ defined above evolves according to
\begin{eqnarray}
i\hbar \frac{\partial}{\partial t} \psi_{12}({\bf x}_1,{\bf
x}_2,t)=(-\frac{\hbar^2}{2m}(\nabla_1^2+\nabla_2^2)+\tilde{V}({\bf
x}_1)+\tilde{V}({\bf x}_2)\nonumber\\ +\tilde{U}_{12}
\delta^{(3)}({\bf x}_1-{\bf x}_2)) \psi_{12}({\bf x}_1,{\bf
x}_2,t), \label{psi12}
\end{eqnarray}
with the initial condition $\psi_{12}({\bf x}_1,{\bf
x}_2,0)=\phi_0({\bf x}_1)\phi_0({\bf x}_2)$. We assume a
spherically symmetric harmonic potential $\tilde{V}({\bf
x})=\frac{1}{2}m\tilde{\omega}^2 {\bf x}^2$, implying
$\phi_0({\bf
x})=(\frac{m\tilde{\omega}}{\pi\hbar})^{\frac{3}{2}}e^{-\frac{m\tilde{\omega}{\bf
x}^2}{2\hbar}}$.

It is convenient to transform to center-of-mass and
relative coordinates defined by ${\bf X}=\frac{{\bf
x}_1+{\bf x}_2}{\sqrt{2}}$ and ${\bf r}=\frac{{\bf
x}_1-{\bf x}_2}{\sqrt{2}}$. In these coordinates the wave
function is separable at all times, $\psi_{12}({\bf X},{\bf
r},t)=e^{-i\frac{\tilde{\omega}}{2}t}\phi_0({\bf X})
\psi({\bf r},t)$. The center of mass wave function exactly
remains in the ground state of $\tilde{V}$. The relative
coordinate wave function $\psi({\bf r},t)$ fulfills the
equation
\begin{equation}
i \frac{\partial}{\partial t}\psi({\bf
r},t)=(-\frac{\hbar^2}{2m}\nabla^2+\tilde{V}({\bf
r})+\bar{U}_{12} \delta^{(3)}({\bf r})) \psi({\bf r},t),
\end{equation}
where $\bar{U}_{12}=\tilde{U}_{12} 2^{-\frac{3}{2}}$. The
interaction between the two spin wave excitations inside
the BEC is thus reduced to a fairly simple problem in
one-particle quantum mechanics.

In practice the interaction energy associated with the
$\bar{U}_{12}$ term is two to three orders of magnitude
smaller than the harmonic oscillator energy scale $\hbar
\tilde{\omega}$. As a consequence, the use of perturbation
theory is well justified. Due to the large separation
between the two energy scales, $\psi({\bf r},t)$ remains
essentially proportional to the ground state, see below. However,
there is an energy shift relative to the ground state
energy, which is given by
\begin{equation}
\Delta E=\langle \phi_0| \bar{U}_{12} \delta^{(3)}({\bf
r})) |\phi_0\rangle=\bar{U}_{12} |\phi_0({\bf
0})|^2=\bar{U}_{12} s^{-3}, \label{shift}
\end{equation}
where $s=\sqrt{\frac{\pi\hbar}{m\tilde{\omega}}}$ is the
characteristic length scale of the ground state wave
function, which is related to its full width at half
maximum $l$ by $s=\sqrt{\frac{\pi}{8\ln 2}}l$. This energy
shift is the basis of our quantum gate proposal. Since it
is due to the interaction, it only occurs if there are two
excitations in the condensate, allowing one to realize a
controlled phase gate. The gate naturally has high fidelity
\cite{he} because the correction terms to the ground state
wave function have amplitudes of order $\frac{\Delta
E}{\hbar \tilde{\omega}} \sim \frac{a_{12}}{s}$, which is
smaller than $10^{-2}$ even for the largest scattering
length and smallest ground state size that we will
consider. This means that, apart from the phase, the
overlap with the initial state remains extremely high,
which is exactly what is required for high-fidelity
operation \cite{he}.

The remaining challenge is therefore to achieve a
controlled phase shift of $\pi$. Let us begin by choosing
parameter values that should be straightforwardly
achievable. For example, one can choose level 0 in the
$F=1$ sub-manifold of $^{87}$Rb, and levels 1 and 2 in the
$F=2$ sub-manifold, giving $a_{00}=5.39$ nm,
$a_{01}=a_{02}=5.24$ nm and $a_{12}=5.58$ nm
\cite{duttonthesis}, and a full width at half maximum for
the ground state wave packet $l=8 \mu$m (corresponding to
about ten wavelengths). With these values one finds that
the time required for a phase of $\pi$ is 6 minutes, which
at first sight may seem rather discouraging. We now discuss
how to overcome this difficulty by acting both on the
$\bar{U}_{12}$ factor and the $s^{-3}$ factor in Eq.
(\ref{shift}).

The factor
$\bar{U}_{12}=\frac{\sqrt{2}\pi\hbar^2}{m}(a_{12}-\frac{a_{01}
a_{02}}{a_{00}})$ is very small for the values given above
because there is a quasi-cancelation between the two terms
in parentheses because all the scattering lengths are so
similar. A moderate increase in $a_{12}$, which can be
achieved using a Feshbach resonance
\cite{inouye,volz,kaufman}, can therefore lead to a very
large increase of $\bar{U}_{12}$. For example, increasing
$a_{12}$ by a factor of $F=3$, which was already
demonstrated in Ref. \cite{volz} for $^{87}$Rb, increases
$\bar{U}_{12}$ by a factor of 24.

\begin{figure}
\epsfig{file=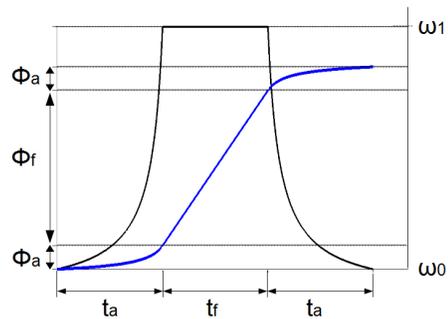,width=0.7 \columnwidth}
\caption{Qualitative time dependence of the trap frequency
$\omega$ and of the acquired phase $\phi$. The total time
is $2 t_a+t_f$, where $t_a$ is the time required for
adiabatically changing the trap frequency from $\omega_0$
to $\omega_1$ or vice versa, and $t_f$ is the time during
which the trap frequency is held fixed at the high value
$\omega_1$. The corresponding phases are $\phi_a$ and
$\phi_f$, giving a total phase $2\phi_a+\phi_f$, where
$\phi_f \gg \phi_a$ in the discussed regime.}
\label{phiomega}
\end{figure}

A comparable gain can be achieved by acting on the second
factor in Eq. (\ref{shift}), i.e. on the size of the wave
function, or equivalently on the trapping frequency. We
already mentioned in the introduction that $l$ (and thus
$s$) cannot be too small during the storage process,
because focusing becomes too difficult and the slowly
varying envelope approximation breaks down. However, the
trapping frequency can be increased once the photons have
been stored, see Fig. 2, with the caveat that this increase has to be done adiabatically so that the spin waves remain in the
ground state of the effective trapping potential. The
mentioned $l=8 \mu$m corresponds to an effective frequency
$\tilde{\omega}=2\pi$ 10 Hz, which corresponds to a real
trap frequency $\omega=2\pi$ 50 Hz. This gives a condensate size of 17 $\mu$m for $N=10^5$ atoms in the Thomas-Fermi approximation \cite{DalfovoRMP}. The effective frequency can be increased
to $\tilde{\omega}=2\pi$80 Hz in $t_a=0.14$ seconds while
exciting the system out of the ground state with a
probability that is smaller than $0.002$. At this frequency
the ground state size $l$ is $2.9 \mu$m and the size of the condensate is 7.4 $\mu$m. For a Feshbach
factor $F=3$ a phase of order $\pi$ can then be achieved
with $t_f=0.73$ seconds. Taking into account that one has to
decrease the frequency before readout, the total gate time
$2 t_a+t_f$ is 1.01 seconds for this example. Note that there is
also a small contribution to the total phase from the
adiabatic compression and expansion periods, see Fig.
\ref{phiomega}. The peak density of the condensate in its compressed state is $6 \times 10^{14}/$cm$^3$ in this case, which is compatible with typical three-body loss rates \cite{loss}.
Shorter gate times could be achieved for smaller initial ground state sizes, higher compressed densities, or larger Feshbach enhancement factors.

We have shown that a controlled phase of $\pi$ between
individual photons is achievable on the one-second timescale under realistic conditions. We hope that our
proposal will stimulate experimental work in this
direction.

We thank D. Feder, A.I. Lvovsky, A. MacRae and A. S{\o}rensen for
helpful comments. This work was supported by an AI-TF New Faculty Award and an NSERC Discovery Grant.


\begin{thebibliography}{99}
\bibitem{phasesep} These conditions are compatible
with the phase-separation regime used in Ref. \cite{zhang},
provided that $a_{11},a_{22} < \frac{a_{01}^2}{a_{00}}$.
\end{thebibliography}
\end{document}